\documentclass[12pt,a4paper,final]{iopart}

\usepackage{iopams}
\usepackage{graphicx}
\usepackage[breaklinks=true,colorlinks=true,linkcolor=blue,urlcolor=blue,citecolor=blue]{hyperref}
\usepackage{cite}

\begin{document}
\title[J Guthrie and J Roberts]{A scalable theoretical mean-field model for the electron component of an ultracold neutral plasma}

\author{John Guthrie and Jacob Roberts}
\address{Department of Physics, Colorado State University, Fort Collins, CO 80523, USA}
\ead{john.guthrie@colostate.edu}

\begin{abstract}
The electron component of an ultracold neutral plasma (UCP) is modeled based on a scalable method using a self-consistently determined mean-field approximation. Representative sampling of discrete electrons within the UCP are used to project the electron spatial distribution onto an expansion of orthogonal basis functions. A collision operator acting on the sample electrons is employed in order to drive the distribution toward thermal equilibrium. These equilibrium distributions can be determined for non-zero electron temperatures even in the presence of spherical symmetry-breaking applied electric fields. This is useful for predicting key macroscopic UCP parameters, such as the depth of the electrons' confining potential. Dynamics such as electron oscillations in UCPs with non-uniform density distributions can also be treated by this model.

\end{abstract}

\vspace{2pc}

\section{Introduction}
Since their creation ultracold neutral plasmas (UCPs) \cite{rolston99,pillet00,dutta01} have provided a rich system of studying plasma physics in a unique parameter space, including influences of strong coupling \cite{kulin00,killian01,murillo01,ichimaru82,bannasch12,lyon13}. Determining UCP properties such as electron temperature and internal electric fields \textit{in situ} is typically a difficult task due to the fact that physical probes (e.g. Langmuir probes \cite{lam65}) placed inside the plasma would destroy the UCP. Many UCP experiments, especially those focused on the electron component, utilize detection schemes that rely on particle escape and extraction with the assistance of applied external fields in order to address these problems \cite{roberts04,pillet05}. In most cases extrapolating information about the plasma from this extraction process depends on accurately knowing properties such as the depth of the confining potential. In addition, it is useful to calculate electron dynamics such as electron plasma frequencies in order to determine parameters like the UCP density, for example \cite{wilson13-2}. Fortunately, UCPs also provide a relatively clean environment apt for computational modeling. However, the long-range nature of the Coulomb forces that dominate UCP dynamics makes simulation a nontrivial task.

A number of different simulation schemes have been previously utilized for modeling ultracold plasmas using few approximations \cite{robicheaux02,kuzmin02-1,pohl04,murillo06,kuzmin02-2,christlieb06,jeon08,bannasch11}. The most complete models are molecular dynamics (MD) simulations that treat the plasma as a collection of $N$ individual particles interacting via Coulomb forces\cite{robicheaux02,kuzmin02-1,pohl04,murillo06,kuzmin02-2}. These methods face difficulties due to the long-range nature of the Coulomb force in addition to other effects such as disparate timescales between short orbit electrons bound to an ion and unbounded electrons. The long-range Coulomb forces that drive the dynamics of the individual particles make for $\Or(N^2)$ calculations in the most natural implementation of a numerical model. This makes MD simulations computationally expensive thereby limiting their usefulness when studying large numbers of particles or long time scale effects \cite{kuzmin02-2}. 

Other methods have been developed that are designed to reduce the $\Or(N^2)$ complexity of these simulations by making various approximations. Methods that adaptively subdivide the space around each particle, such as the TREE model \cite{jeon08}, are able to reduce the scaling to $\Or(N\log N)$ \cite{barnes86}. The state of the art Fast Multipole Method \cite{bannasch11} that approximates batches of distant charges as multipoles rather than individual sources can be shown to reduce the calculations to $\Or(N)$ \cite{greengard87}. While these approximations can provide increased efficiency for sufficiently large $N$ there is non-trivial computational overhead in implementing these methods, and so simulations can still often be computationally expensive. While these methods can provide a fairly complete description of the plasma, less computationally intensive techniques are still useful.

In principle, one way of reducing the complexity would be to decrease the number of particles that require calculations below the total $N$ in the system being modeled. However, this reduction cannot be as simple as reducing the number of particles by some factor and increasing their charge and/or mass by an appropriate factor as this leads to inconsistent scaling of characteristic plasma parameters. That is, if the number of particles is changed there is no way to simultaneously preserve relationships between, for instance, the spatial electron distribution and the electron-electron collision rate through adjusting charge and mass values. We have included an Appendix that details these scaling problems at the end of this article.

Instead, the model we developed details a scalable simulation suitable for extracting macroscopic UCP parameters. This is accomplished by averaging the electric field inside the plasma using a mean-field approximation. The mean field is calculated self-consistently using Monte Carlo sampling of individual electrons that move under the influence of the mean field. These $S_\rme$ sample electrons can be used as a representation of the distribution of the total $N_\rme$ electrons being modeled in the system by scaling their statistical weight while still maintaining their standard electron charge and mass. By constructing a suitable set of basis functions, the sample electrons can be projected onto a series of these functions in order to generate a smoothed approximation of the discrete distribution. This lets us compute forces on the sample electrons with scaling $\Or(S_\rme)$ which gives us a scalable way of determining individual electron trajectories as the plasma evolves. Momenta and energy exchange events can be applied to these sample electrons to drive them to a thermal distribution.

Macroscopic parameters can be extracted from both the discrete sample electrons as well as the smooth, approximate distribution. As an example and test of the technique, we examine the convergence of the potential depth as a function of the number of basis expansion terms used in the calculation. Additionally, we will present simulation results for the potential depth of the plasma as a function of applied DC electric field strength. Finally, we will explore extensions for this model such as the analysis of the dynamics of the electron component when subjected to applied electric field pulses that produce electron plasma oscillations.

\begin{figure}[htb!]
  \centering
  \includegraphics*{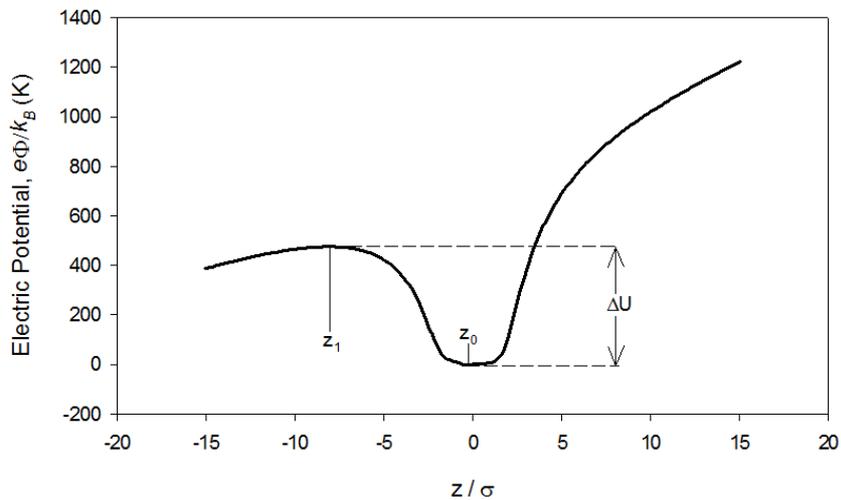}
  \caption{\label{figure1}A typical electric potential curve, in scaled temperature units, along the z axis for an ultracold plasma under the influence of a DC electric field pointed along the z axis. The ion density distribution is assumed to be a spherically symmetric Gaussian $\propto \exp\left(-r^2/2\sigma^2\right)$. The positions of the local extrema in the curve are marked by $z_0$ and $z_1$. The potential depth of the plasma, $\Delta U$, is also illustrated. This was taken from data using $\sigma$ = 800 $\mu$m; $T = 20$ K; $E_{\mathrm{DC}} = 3$ V/m; ion number $N_\rmi$ = 200,000; and electron number $N_\rme$ = 110,000.}
 \end{figure}

As mentioned, one purpose of the calculations presented in this paper is to extract the predicted potential depth of a UCP from a fast, approximate computational model. Various experiments focused on studying the electron component of UCPs have looked toward electron evaporation and escape as measurable quantities from which additional information about the plasma may be extracted. For instance, the depth parameter is useful for determining a predicted electron evaporation rate that can be compared to experimentally measured rates \cite{wilson13-1}. The parameter is also relevant for certain electron temperature measurement techniques that rely on electron extraction from the plasma using applied electric fields \cite{roberts04}. The plasma potential can be calculated with much less difficulty by using systems with simplifying assumptions such as spherical symmetry \cite{robicheaux03,collins08} or zero temperature electron distributions \cite{rolston10}. However, in many experiments a spherical symmetry-breaking external electric field is an integral part of the system as a means for guiding electrons for detection. Figure \ref{figure1} shows a typical potential energy curve for an electron, computed by our model, along the z axis of a UCP under the influence of a DC electric field. Breaking spherical symmetry will often increase the complexity of modeling certain systems by changing calculations from solving ordinary differential equations to partial differential equations, for example.  Finite temperature effects will also play a role in shaping the electron thermal distribution and ultimately affect the depth parameter. Factors such as these have motivated us to find a more complete computational model for efficiently calculating plasma parameters such as potential depth.

In addition to equilibrium parameters such as the potential depth, we are also interested in characterizing some dynamics of the electrons. From the classical plasma definitions for the two-body collision rate $\nu_{ee} \sim \sqrt{n}~\Gamma^{3/2} \ln\left(\alpha \Gamma^{-1}\right)$ \cite{spitzer} and electron plasma oscillation frequency $\omega_p \sim \sqrt{n}$ we find that the dimensionless ratio $\nu_{ee} / \omega_p$ scales as $\Gamma ^{3/2}  \ln\left(\alpha \Gamma^{-1}\right)$,  where $n$ is the electron density, $\alpha$ is a constant, and $\Gamma$ is the Coulomb coupling parameter \cite{ichimaru82}. For $\Gamma \ll 1$ when the plasma is weakly coupled, as is the case for the UCPs modeled in this paper, the collision rate is much less than the oscillation frequency. Therefore the electron dynamics can be treated using a mean-field approximation to a reasonable degree of accuracy.

In this paper we make a number of assumptions about the structure of the UCP. In typical experiments, the UCP is formed by laser ionizing a cloud of neutral atoms \cite{rolston99}. We assume the atoms have a Gaussian spatial density distribution, and after ionization the ions retain the same structure. The frequency of the photoionizing laser can be tuned above the ionization threshold and therefore provides control over the initial kinetic energy of the electrons. The resulting cloud of ions and electrons is initially neutral and allows a fraction of electrons to escape; we assume this escape fraction can be controlled experimentally, e.g. through the use of an applied electric field. The electrons that remain inside the plasma will then experience space charge confinement due to the resulting charge imbalance. At this point the confined electrons will exert a thermal pressure on the ions and drive expansion of the ion cloud \cite{kulin00}. While this expansion is not included in the model detailed in this paper, it could be included by extension in future work. For practical purposes, UCP experiments focused on measuring the electron component use electric fields as a means of controlling and guiding electron escape \cite{roberts04,rolston10,raithel08}. Our model includes an applied DC electric field, and this is indeed one of the motivating factors for developing these models. It does not incorporate an applied magnetic field at this time, however extending the model to include the effects of magnetic fields will be the subject of future work.

\section{Basis function expansion}

The crux of the method presented in this paper is the approximation of the electron density, $n_{\rme}(\mathbf{x})$, using an expansion of suitable basis functions. The model presented here incorporates a DC electric field along the z-axis which breaks the spherical symmetry but maintains axial symmetry. We therefore assume a separable, axially symmetric form for the electron density given by
\begin{equation}
n_{\rme}(r, \theta, \phi) = N_\rme\sum_{l,n}a_{l,n}R_{l,n}(r)\,Y_{l}^{0}(\theta,\phi)\label{nseries}
\end{equation}
where the angular part is specified by the spherical harmonics, $Y_l^m(\theta,\phi)$ with $m=0$ due to symmetry, and the $a_{l,n}$'s are the expansion coefficients. The radial functions $R_{l,n}(r)$ are constructed from polynomials modified by a Gaussian:
\begin{equation}
R_{l,n}(r) = f_{l,n}\!\left(\frac{r}{\sigma}\right)\exp\left(\!-\frac{r^2}{2\sigma^2}\right).\label{rln}
\end{equation}
The $f_{l,n}$ functions are dimensionless polynomials calculated using Gram-Schmidt orthogonalization. These polynomials are chosen so that the radial functions satisfy the orthogonality relation 
\begin{equation}
\int_0^\infty\!R_{l,n}(r)\,R_{l,n'}(r)\,r^2\,\mathrm{d}r = \sigma^3 q_{l,n} \delta_{nn'}\label{rorth}.
\end{equation}
The $q_{l,n}$ factor on the right hand side is a numerical constant determined by enforcing two different normalization conditions. For $l\neq 0$ the constant is equal to 1, but for all $l = 0$ terms\textemdash which we recognize as the monopole part of the distribution\textemdash it is a numerical factor that results from each individual function being normalized over all space to unity. In order to avoid singularities at the origin when calculating the electric potential from these functions we construct a different basis set for each $l$ that begins with the term $(r/\sigma)^l$, as is typical. For example, the $l = 0$ set comes from orthogonalizing the terms $\lbrace 1, \frac{r}{\sigma}, \left(\frac{r}{\sigma}\right)^2\!, ... \rbrace $ whereas the $l = 1$ set is derived from $\lbrace \frac{r}{\sigma}, \left(\frac{r}{\sigma}\right)^2\!, \left(\frac{r}{\sigma}\right)^3\!, ...\rbrace$.

We can now use (\ref{rorth}) to project the electron density, (\ref{nseries}), onto each basis function and retrieve the corresponding $a_{l,n}$ coefficient:
\begin{equation}
a_{l,n} = \frac{1}{N_\rme\,\sigma^3\,q_{l,n}}\int\!\mathrm{d}\Omega\int_0^\infty\!r^2\,\mathrm{d}r\,n_\rme(r,\theta,\phi)\,R_{l,n}(r)\,Y_{l}^{0}(\theta,\phi). \label{alns}
\end{equation}
Once each of these coefficients is known it is possible to construct the approximated, smoothed electron density function.

If we wish to keep the number of basis functions required to accurately approximate the electron distribution reasonably low then we must address the issue of projecting smooth functions on a finite spatial size distribution. The finite nature of the distribution will lead to a boundary at which the electron density will become exactly zero outside, and without enough basis functions the smoothed distribution will have difficulties accurately representing this boundary. In most cases we found that the basis functions produced a distribution that would go negative or have a local minimum at this boundary. We observed that the unphysical non-zero portions of the distribution outside of the boundary could generate anomalous electric fields inside of the plasma large enough to affect the electrons' dynamics.

Our solution to this issue is to approximate the boundary as a sphere around the electron center of mass and redefine the basis functions to be piecewise with a cut-off radius beyond which they are exactly zero. This cut-off radius is found by searching the projected distribution for the zeros and local minima mentioned above. It is these cut-off, piecewise basis functions that are used to calculate the electric field and potential for the electron component of the plasma. Fortunately, this piecewise cut-off did not prevent us from analytically determining the functions for the electric potential and fields.

Using cut-off distributions to calculate the electric fields implies that electrons that cross the boundary and escape the plasma do not have an effect on the remaining electrons. This is a limitation of this model, but we don't expect it to play a significant role for the conditions that we treat in this article which correspond to realistic UCP experimental parameters well after formation. The plasmas we studied here have a small escape fraction. Furthermore the applied DC electric field quickly carries the escaped electrons away, and they therefore have a negligible impact on the UCP.

The electric potential $\Phi_{\rme}(\mathbf{x})$ generated from each basis function can be calculated by inserting the series approximation (\ref{nseries})\textemdash with $R_{l,n}(r)$ now being the piecewise cut-off functions\textemdash into the integral form of Poisson's equation with $\rho_{\rme}(\mathbf{x}) = -e\,n_{\rme}(\mathbf{x})$, where $e$ is the elementary charge. We can take advantage of the fact that the angular part of $\rho_{\rme}$ can be expressed in spherical harmonics by replacing the Green's function $1/|\mathbf{x} - \mathbf{x'}|$ in Poisson's equation with its Laplace expansion \cite{arfken}. This allows the angular part of the integral to be computed using the orthogonality condition for $Y_l^m(\theta,\phi)$. The potential can now be expressed as the following series of radial integrals:
\begin{equation}
\Phi_{\rme}(\mathbf{x}) = -\frac{e N_{\rme}}{\epsilon_0}\sum\limits_{n=0}^{\infty}\sum\limits_{l=0}^{\infty}\frac{a_{l,n}}{2l+1}\left[r^{-(l+1)}A_{l,n}(r) + r^lB_{l,n}(r)\right]Y_l^0(\theta,\phi)\label{phiseries}
\end{equation}
where
\begin{equation*}
A_{l,n}(r) = \int\limits_{0}^{r}\!R_{l,n}(r')\,{r'}^{l+2}\,\mathrm{d}r' \quad \mathrm{and} \quad B_{l,n}(r) = \int\limits_{r}^{\infty}\!R_{l,n}(r')\,{r'}^{1-l}\,\mathrm{d}r',
\end{equation*}
and $\epsilon_0$ is the vacuum permittivity. Since each $R_{l,n}(r')$ is simply a series of polynomial-Gaussian products inside the cut-off and constant zero outside we can analytically compute every above term. The resulting expressions for the $A_{l,n}$ and $B_{l,n}$ terms consist of sums of error functions and polynomial-Gaussian products.

The objective of this method is to have the ability to calculate the trajectories of electrons in the plasma using a mean-field approximation.  The electrons' contribution to the net electric field $\mathbf{E}_{\mathrm{net}}$ can be calculated from (\ref{phiseries}) using $\mathbf{E} = -\vec{\nabla}\Phi$:
\begin{equation}
\eqalign{\fl\quad\mathbf{E}_\rme\!\cdot\hat{r} = \frac{e N_{\rme}}{\epsilon_0}\sum\limits_{n=0}^{\infty}\sum\limits_{l=0}^{\infty}\frac{a_{l,n}}{2l+1}\left[\frac{-(l+1)}{r^{l+2}}A_{l,n}(r) + l\,r^{l-1}B_{l,n}(r)\right]Y_l^0(\theta,\phi) \cr
\fl\quad\mathbf{E}_\rme\!\cdot\hat{\theta} = \frac{e N_{\rme}}{\epsilon_0}\sum\limits_{n=0}^{\infty}\sum\limits_{l=0}^{\infty}\frac{a_{l,n}}{2l+1}\left[{r^{-(l+2)}}A_{l,n}(r) + r^{l-1}B_{l,n}(r)\right]\left[\!\sqrt{l(l+1)}\,\rme^{-\rmi \phi}\,Y_l^1(\theta,\phi)\right]} \label{Evec}
\end{equation}

An identical treatment can be performed for the ion density $n_\rmi(\mathbf{x})$. The ions are not expected to evolve as rapidly as the electrons since they are much more massive. In this article, we focus on short timescale electron motion such that the ion component of the UCP does not change significantly and could thus be treated as constant. Our method of the mean-field treatment of the electrons could be extended to the ions to capture their dynamics, if desired. We make the approximation that shortly after ionization the ions remain as a spherically symmetric Gaussian distribution fixed at the origin,
\begin{equation*}
n_\rmi(r) = \frac{N_\rmi}{{(2\pi\sigma^2)}^{3/2}}\,\exp\left(\!-\frac{r^2}{2\sigma^2}\right),
\end{equation*}
with the corresponding electric potential and field
\begin{equation}
\eqalign{\Phi_{\rmi}(r) = \frac{e}{4 \pi \epsilon_0} \frac{N_{\rmi}}{r}\, \mathrm{erf} \left(\frac{r}{\sqrt{2} \sigma}\right)\cr
\mathbf{E}_{\,\rmi}(r) = \frac{e}{4 \pi \epsilon_0} \frac{N_{\rmi}}{r^2} \left[\mathrm{erf}\left(\frac{r}{\sqrt{2}\sigma}\right) - \sqrt{\frac{2}{\pi}}\frac{r}{\sigma}\, \exp \left(\!-\frac{r^2}{2\sigma^2}\right) \right]. }\label{ni}
\end{equation}

With the total mean electric field we calculated trajectories of individual electrons within the plasma using Newtonian mechanics. Starting from some spatial distribution of discrete electrons we projected that distribution onto each density basis function to determine the $a_{l,n}$ coefficients. Once the coefficients were found they could be used to determine the mean electric field from the electron component of the plasma by utilizing (\ref{Evec}). Combining the field from the electrons with the ion component's field in (\ref{ni}) and the applied DC field let us compute the force on each electron, $\mathbf{F}=-e\left(\mathbf{E}_{\rme} + \mathbf{E}_{\,\rmi} + \mathbf{E}_{\mathrm{DC}}\right)$. Calculating individual electron trajectories from these forces allows the system to evolve self-consistently.

Since the discrete electrons are free to move independently under the influence of the calculable mean-field forces we do not need to make any assumptions about the velocity distribution of the particles. Additionally, if a mechanism is present that can mediate exchanges of energy and momentum between the particles then the distribution will approach thermal equilibrium. To create thermal equilibrium distributions easily and efficiently, a non-physical collision operator that randomizes the velocities of nearest-neighbour pairs of electrons is included in our model. Thus a thermal distribution for the electron component under the influence of a DC electric field can ultimately be found. The H-theorem \cite{kubo} guarantees that the same thermal equilibrium state will be achieved regardless of the details of the collision operator as long as certain usual assumptions (e.g. time-reversibility) are observed.

\section{Simulation details}
One of the advantages of this mean field approximation is that we are not required to account for every individual particle in the plasma. Instead it is possible to find the basis expansion coefficients and evaluate certain macroscopic parameters using only a statistical sampling of electrons. We do \textit{not} scale the charge nor mass of these sample electrons as they are not single particle representations of multi-electron bunches. When the forces from the mean-field approximation act on the sample electrons they are treated as having the standard electron charge and mass. What is being scaled here instead is a statistical weighting parameter that accounts for the fact that we're describing the distribution of the total $N_\rme$ electrons using a fewer number of sample particles. We choose some sample electron number, $S_\rme < N_\rme$, and use their positions as a representation of the entire electron density,
\begin{equation}
n_\rme(\mathbf{x}) \approx \frac{N_\rme}{S_\rme}\sum\limits_{i = 1}^{S_\rme}\delta(\mathbf{x} - \mathbf{x}_i)\label{repdensity}.
\end{equation}
Inserting (\ref{repdensity}) into (\ref{alns}) gives us an expression for determining the expansion coefficients from our representative sample,
\begin{equation}
a_{l,n} = \frac{1}{S_\rme\,\sigma^3\,q_{l,n}}\sum\limits_{i = 1}^{S_\rme}R_{l,n}(r_i)\,Y_l^0(\theta_i,\phi_i).
\end{equation}

Using these coefficients and our expansion for the mean electric field we found the trajectories of our sample electrons by numerically integrating the second order ODE $\ddot{\mathbf{x}} = -e\,\mathbf{E}(\mathbf{x})/m_\rme$ where $m_\rme$ is the electron mass. We would typically calculate trajectories over durations on the order of 300 $\omega^{-1}$ and recalculate the coefficients every 0.3 $\omega^{-1}$ or so, where $\omega^{-1} = \left(n_{\mathrm{i ,peak}} e^2 / m_\rme \epsilon_0\right)^{-1/2}$ is the natural timescale for our system and $n_{\mathrm{i, peak}} = N_\rmi/\left(2\pi\sigma^2\right)^{3/2}$. As the electrons move the spatial distribution changes and thus the $a_{l,n}$ coefficients change as well. In our implementation of the calculations we use a predictor-corrector method with self-consistency checks so that the coefficients may evolve continuously throughout the integration of the trajectories.

As mentioned above, we do not need to make any assumptions about the velocity distribution of the electrons. Because of the finite nature of the plasma's potential depth we do not expect the speeds of the electrons to be described by a Maxwell-Boltzmann distribution as the high velocity electrons on the tail of the distribution would be free to leave the plasma. Previous studies have indicated that approximating the electron speeds as a truncated Maxwell-Boltzmann or Michie-King distribution would provide a more accurate description \cite{pillet05,collins08}. However, we need not make any of these assumptions in our implementation of these calculations. 

The sample electrons are initialized to a Maxwell-Boltzmann distribution, but as the plasma evolves pseudo-collisions are used to bring the electrons to thermal equilibrium. A togglable elastic collision operator acts on the electron cloud between integration time steps in such a way that velocities of pairs of electrons are randomized while preserving the kinetic energy and center of mass momentum. A single random electron is chosen and a scan is done to find its nearest neighbour in order to ensure that the difference in potential energy of the electrons is small compared to their kinetic energy. Transforming to the center of mass frame of the electron pair allows us to rotate their relative velocity vector to a random direction while preserving its magnitude. This allows the electrons to change velocity while conserving the total momenta and energy of the pair.

The combination of the integration and collision techniques described above allows for another advantage for this simulation: trajectories of electrons can be integrated individually as opposed to integrating the entire cloud all at once. Interpolating the mean-field coefficients with our predictor-corrector method implies that the behaviour of individual sample electrons does not directly affect the trajectory of any other electron during integration time steps. Furthermore, the pseudo-collisions do not require a spatial overlap of electron paths and only operate between the integrations. This allows us to determine the trajectories of each electron one at a time instead of integrating the entire cloud simultaneously. The advantage here is that the convergence of the numerical integration of the system is no longer limited by the fastest moving electrons. If we were forced to integrate the entire system simultaneously then convergence would depend on the electron that requires the most integration subdivisions. By calculating each path individually the quickly changing electron trajectories can be integrated with however many divisions it requires while allowing the slowly changing electrons to be integrated quickly.

With the sample electron trajectories and expansion coefficients at our disposal we are able to calculate a number of macroscopic parameters that describe the electron cloud and the plasma as a whole. The positions of the electrons can be used to track the center of mass motion of the electron component as well as its rms size. Using the velocities we are able to characterize the total kinetic energy of the electrons, $K_{total}\,$, and determine an approximate temperature from $K_{total} \approx \frac{3}{2} N_\rme\,k_B\,T$, where $k_B$ is the Boltzmann constant. It is also possible to track the electron escape rate from the plasma by monitoring how many electrons cross an imaginary planar or spherical boundary to designate which electrons have been counted by a detector or otherwise have escaped.

As mentioned, one plasma parameter that we are interested in extracting is the total potential depth of the plasma. Using the notation in figure \ref{figure1}, we define the potential depth in ``temperature" units $\Delta U = e\left[\Phi\left(z_1\right) - \Phi\left(z_0\right)\right]/k_B$. From (\ref{phiseries}) and (\ref{ni}) it is possible to determine the net electric potential once the $a_{l,n}$ coefficients are known. Using these tools we can determine the plasma potential depth along the symmetry axis in the presence of an external electric field by locating and evaluating the net potential's extrema.

\section{Simulation results for typical UCP parameters}

\subsection{Convergence and equilibrium characteristics}

\begin{figure}[htb!]
  \centering
  \includegraphics*{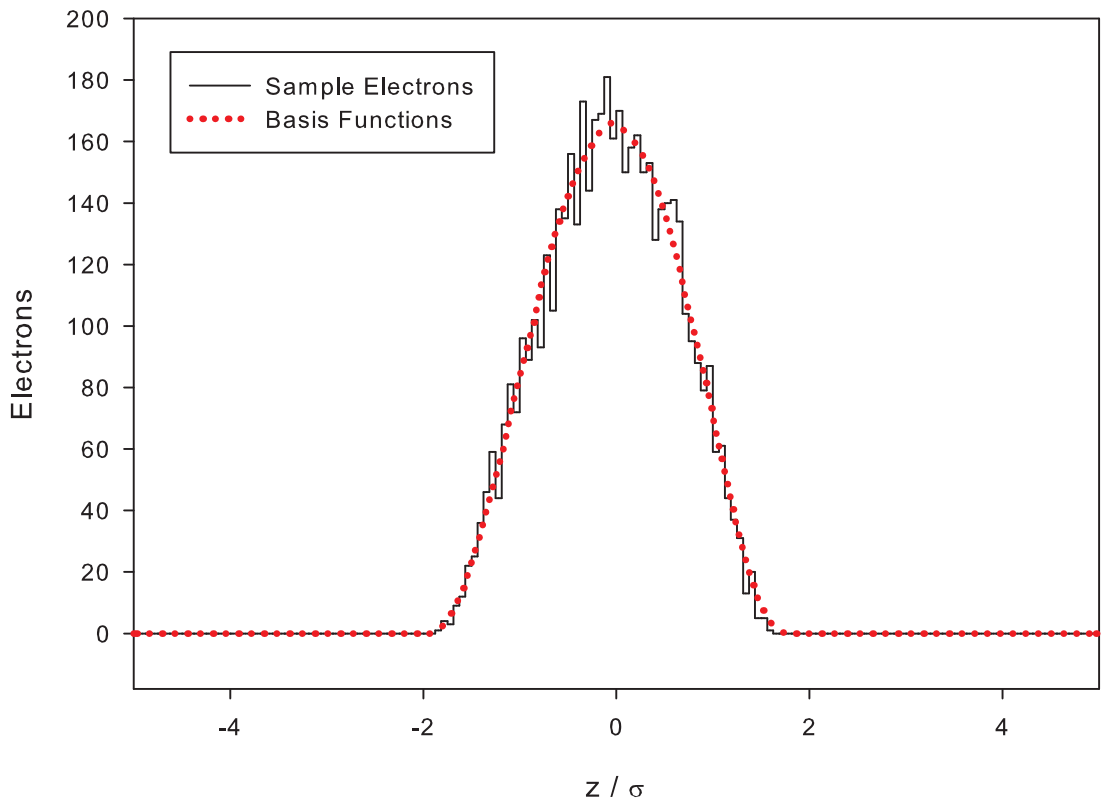}
  \caption{\label{figure2}A typical distribution of the electrons' z coordinates showing good agreement between the discrete electrons and the basis function approximation. The solid black line is a histogram of the individual sample electrons; the dotted red line is the result calculated from the basis functions. This was taken from data with $S_\rme$ = 5,000; $T$ = 5 K; $E_{\mathrm{DC}}$ = 3 V/m; $\delta$ = 0.45.}
 \end{figure}

Our first objective after developing the model was to verify self-consistency between the smoothed basis functions and discrete sample electrons. Figure \ref{figure2} illustrates the good agreement that we found between the spatial distribution calculated from the basis expansion and the distribution of individual sample electrons. The data presented in figure \ref{figure2} is taken from a typical simulation and represents a thermal distribution for the electron component in the presence of an applied DC electric field. For each simulation we are able to tune a number of parameters including the sample electron number $S_\rme$, the applied DC field strength $E_{\mathrm{DC}}$, the electron temperature $T$, the charge imbalance $\delta = (N_\rmi - N_\rme)/N_\rmi$, and the plasma size scale factor $\sigma$. For all of the data presented in this paper we chose $\sigma$ = 800 $\mu$m and $N_\rmi$ = 200,000 in correspondence with measurements made for UCPs formed in a typical experiment in our apparatus.

\begin{figure}[htb!]
  \centering
  \includegraphics*{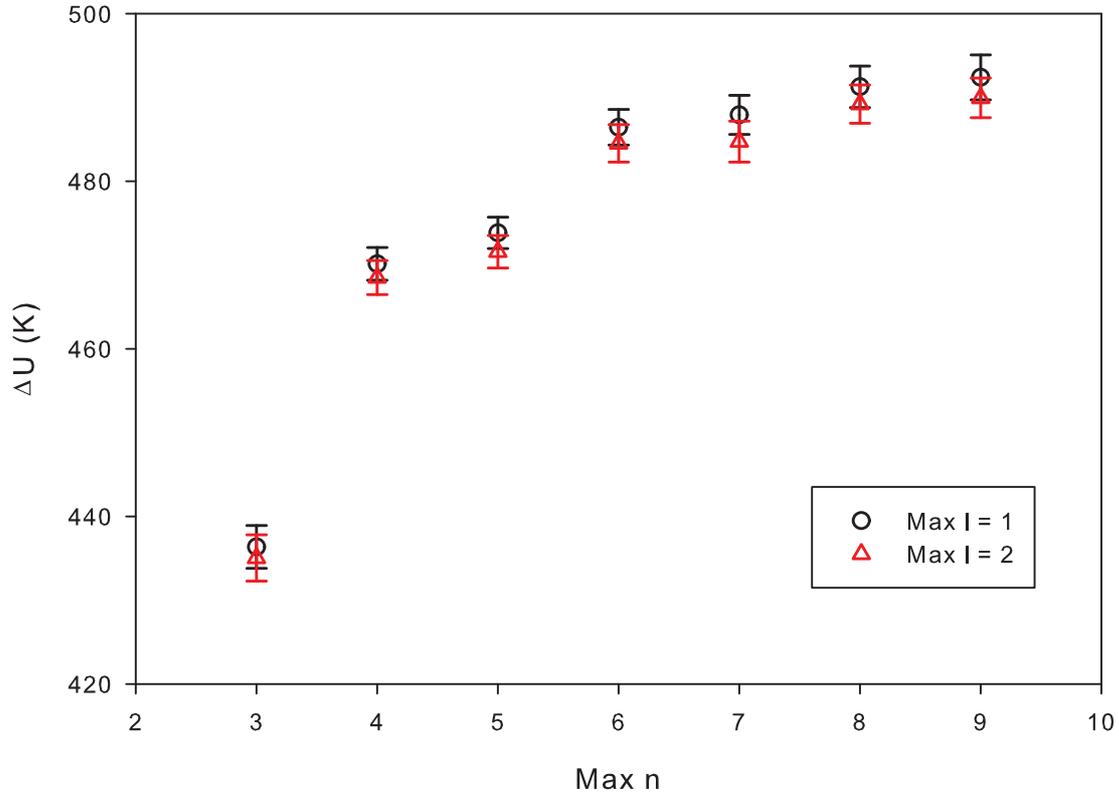}
  \caption{\label{figure3}Demonstration of convergence of the potential depth in temperature units, $\Delta U$, as we increase the number of $n$ terms for each $l$ calculated during a simulation. The black circles are from simulations that utilized $l = \{0, 1\}$ (max $l$ = 1); the red triangles are from $l = \{0, 1, 2\}$ (max $l$ = 2) simulations. This data was taken from simulations at $T$ = 20 K, $E_{\mathrm{DC}}$ = 3 V/m, $\delta = 0.45$.}
 \end{figure}

\begin{figure}[htb!]
  \centering
  \includegraphics*{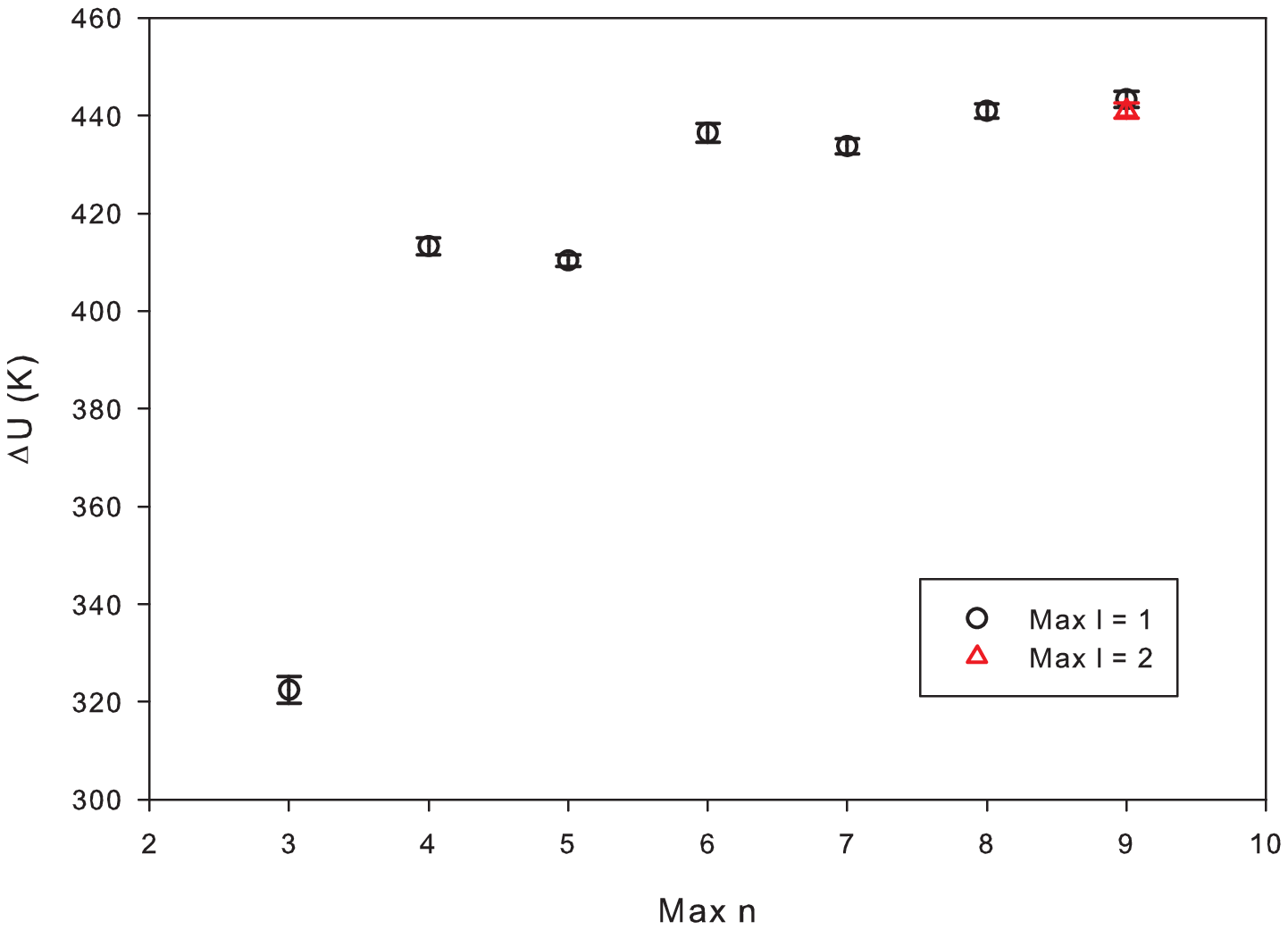}
  \caption{\label{figure4} The same as figure \ref{figure3} except for $T$ = 5 K.}
 \end{figure}

Next we investigated how many terms we needed to calculate in our basis expansion in order for the distribution to converge. This was accomplished by examining macroscopic parameters, in particular the potential depth, extracted from the simulations as a function of the number of terms used. We looked at this by varying the maximum $l$ terms calculated as well as the maximum number of $n$ terms calculated for each $l$ in the expansion. Figure \ref{figure3} shows the convergence of the potential depth in temperature units, $\Delta U$, as we increase the number of $n$ terms for each $l$ in simulations that utilized $l = \{0, 1\}$ (max $l$ = 1) or $l = \{0, 1, 2\}$ (max $l$ = 2). We see that by max $n$ = 6 for max $l$ = 1 the potential depth appears to have converged to within a few percent of the final value. Including $l = 2$ terms in our model adds only small corrections to the data which indicates that higher order terms are not required to get an accurate description. The data was taken by time averaging the potential depth after equilibrium had been established for 12 different simulations each using $S_\rme$ = 5,000; $T$ = 20 K; $E_{\mathrm{DC}}$ = 3 V/m; $\delta$ = 0.45. Figure \ref{figure4} shows data taken using the same method and parameters except at $T$ = 5 K. We see that even at a lower temperature the depth converges with about as many basis functions, and higher $l$ corrections are small as before. On figure \ref{figure4}, this is shown with the max $n = 9$, max $l = 2$ point. With the self-consistency and convergence established we were able to extract plasma characteristics such as the potential depth and examine how they scale with various parameters.

\begin{figure}[htb!]
  \centering
  \includegraphics*{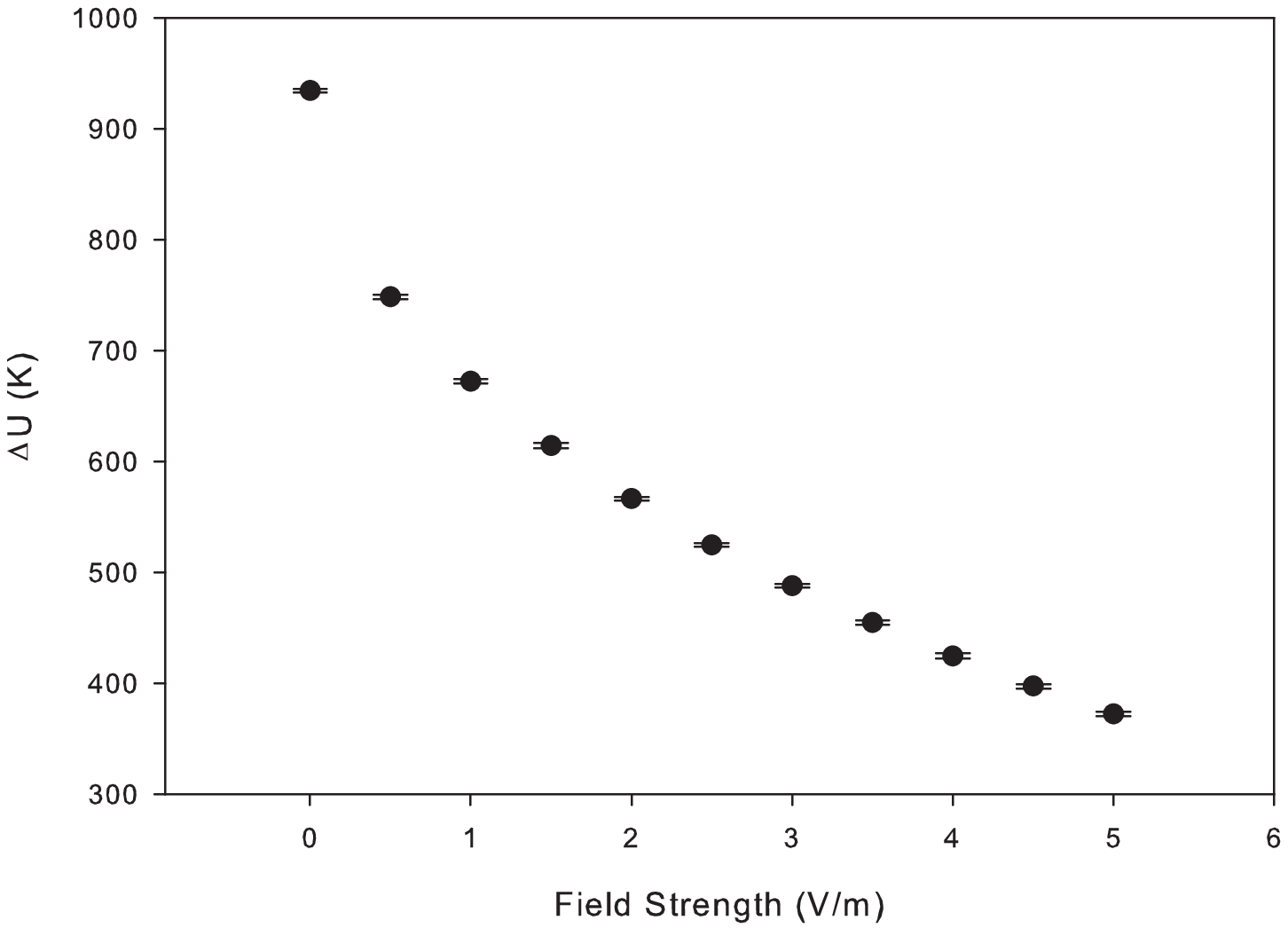}
  \caption{\label{figure5}Plasma potential depth in temperature units as a function of applied DC electric field strength. Taken with parameters $T$ = 20 K, $\delta$ = 0.45.}
 \end{figure}

Typically UCP experiments focused on measuring the electron component utilize electric fields for guiding electrons toward a detector or to tip the confining potential for electron extraction. The strength of the field can be tuned to reduce the depth of the confining potential which in turn can provide information about the electron temperature by detecting the fraction spilled \cite{roberts04,rolston10,raithel08}. Additionally, there is work currently being done examining the possibility of applying forced evaporation in order to cool the electrons \cite{witte14}. Quantifying the dependence of the depth on the field strength is important for calibrating these effects precisely. Figure \ref{figure5} shows our model's prediction for the dependence of the potential depth on the strength of an applied DC electric field with conditions $T$ = 20 K and $\delta$ = 0.45.

\subsection{Electron dynamics after an applied impulse}

\begin{figure}[htb!]
  \centering
  \includegraphics*{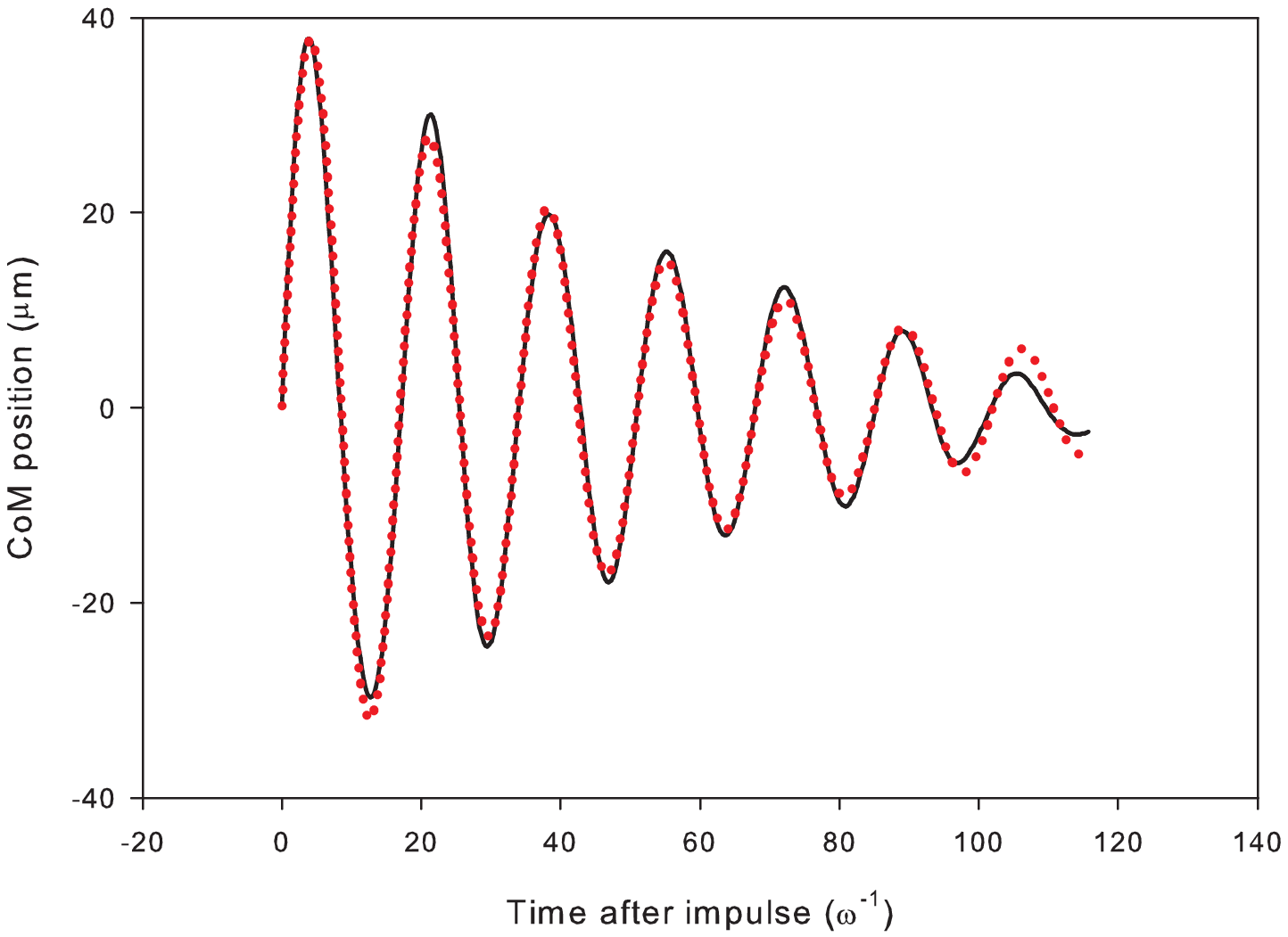}
  \caption{\label{figure6}A typical center of mass trajectory showing electron oscillations and damping after receiving an instantaneous impulse at t = 0. The solid black line is the result from our model; the dotted red line is a fit. Taken with parameters $T$ = 5 K, $E_{\mathrm{DC}}$ = 0 V/m, $\delta$ = 0.45.}
 \end{figure}

\begin{figure}[htb!]
  \centering
  \includegraphics*{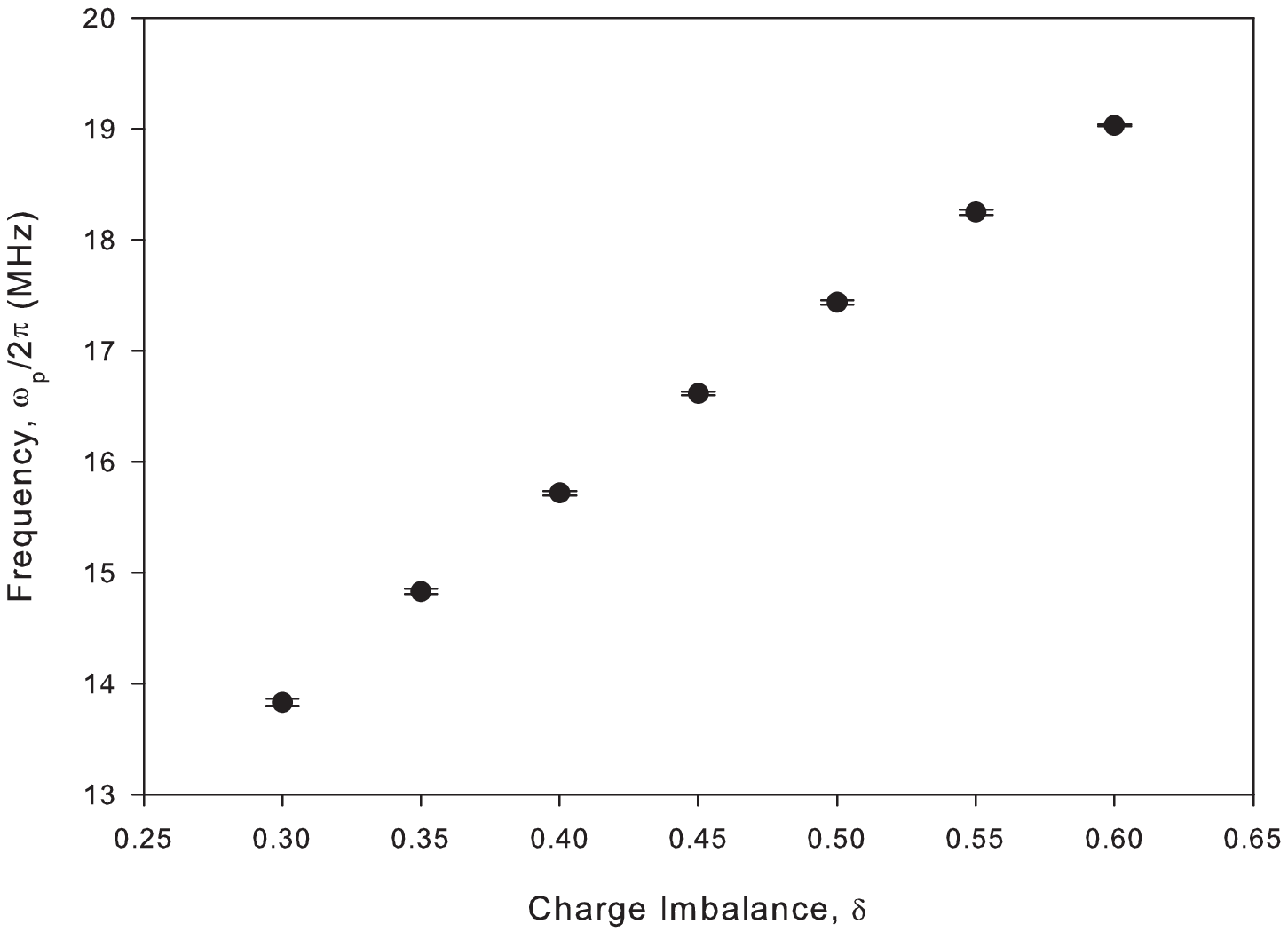}
  \caption{\label{figure7}Electron oscillation frequency calculated from fitting the center of mass motion as a function of charge imbalance. Taken with parameters $T$ = 20 K, $E_{\mathrm{DC}}$ = 0 V/m, $\delta$ = 0.45.}
 \end{figure}

In addition to static properties, this model can be used to study nonequilibrium electron dynamics. Loosely following the experimental design used in reference \cite{wilson13-2}, we applied an instantaneous impulse to the electrons in the z direction by an amount equal to half of the thermal velocity ($\sqrt{k_B T/m_\rme}$). Before the impulse the electrons are initialized to equilibrium using the electron-electron collisions discussed in Section 3. Once the impulse is applied the collisions are toggled off and the electrons undergo oscillations within the confining potential. The center of mass motion of the electrons was fit to a sinusoidal function modified by a decaying exponential as shown in figure \ref{figure6}. From the fit parameters we were able to extract the oscillation frequency of the center of mass. This was modeled for a range of charge imbalances at $T$ = 20 K and $E_{\mathrm{DC}}$ = 0 V/m, the results for which are presented in figure \ref{figure7}. We found that the frequency scales fairly linearly in the $\delta$ range explored roughly agreeing with the results presented in reference \cite{wilson13-2}. There are differences in the calculated oscillation frequency due to the fact that the simplified model of the earlier reference is not strictly correct, and an improved determination of density from our technique is obtained.

As shown in figure \ref{figure6}, our model predicted center of mass damping that fits well with an exponential decay even in the absence of collisions. From the fit we were able to extract the decay time. This non-collisional damping is also observed experimentally. The damping occurs in this case due to the non-uniformity of the density distribution of the ions. This non-uniformity creates anharmonicity in the potential and therefore causes the individual electron's oscillations to dephase. This dephasing was confirmed in simulations we generated using other types of models.

\section{Conclusion}
To summarize, we have developed a scalable method for numerically modeling the electron component of UCPs using a mean-field approximation constructed by projecting a basis function expansion on discrete sample electrons. Our model includes features such as spherical symmetry-breaking DC electric fields and electron-electron thermalizing collisions. Using these tools we have developed a means for determining key UCP characteristics such as the potential depth. This allows us to quantify the effects of an applied electric field on the electrons' confining potential which in turn is useful when studying electron extraction and evaporation from the plasma. We also explored an extension of this model in order to study center of mass dynamics of the electrons after receiving an impulse. This technique allowed us a better determination of UCP density. Future work will include expanding this model to simulate and predict electron escape rates to be compared to experiment and adding the effects of an externally applied magnetic field.

\ack
We acknowledge funding support for this work from the Air Force Office of Scientific Research, grant number FA9550-12-1-0222.

\appendix
\section*{Appendix}
An attractive possibility for modeling UCPs is to run a simulation with a smaller number of electrons than an experimental situation and then scale physical parameters such as charge, spatial extent, mass, etc. in order to maintain the appropriate ratios of e.g. collision rate to plasma oscillation frequency or screening length to plasma spatial extent. In this appendix, we show that such a scaling cannot be done in a way that preserves relevant UCP physical parameter ratios. For instance, one of the defining characteristics of a plasma is its strong coupling parameter $\Gamma = e^2/4\pi\epsilon_0 ak_BT_\rme$, where $a = \left(3/4\pi n\right)^{1/3}$ and $n$ is the density. In this section we assume a Gaussian ion density distribution with spatial size parameter $\sigma$ thus $n \sim N/\sigma^3$. Other defining plasma characteristics are, for example, the two-body collision rate $\nu_{\rme \rme}$, the three-body collision rate $K$, the screening-to-size ratio $\kappa = \lambda_D/\sigma$, and the UCP expansion timescale $t_{\mathrm{exp}}$. We can define the above characteristic parameters using the classical plasma definitions and demonstrate inconsistent scaling by writing them in terms of the following plasma parameters: $N$, $\Gamma$, the ion mass $m_\rmi$, and the plasma frequency $\omega_p$.

\begin{table}
\caption{\label{apptab}Table showing the scaling of the two-body collision rate $\nu_{\rme\rme}$, three-body collision rate $K$, screening-to-size parameter $\kappa$, and expansion timescale $t_{\mathrm{exp}}$.}
\begin{indented}
\item[]\begin{tabular}{@{}lll}
\br
Parameter &Scaling &\\
\mr
$\nu_{\rme\rme}$&$\omega_p\Gamma^{3/2}\ln\left(\alpha\Gamma^{-1}\right)$&\cite{spitzer}\\
$K$&$\omega_p\Gamma^{9/2}$&\cite{killian01}\\
$\kappa$&$N^{-1/3}\Gamma^{-1/2}$&\cite{spitzer}\\
$t_{\mathrm{exp}}$&$\sqrt{m_\rmi}/\omega_p\kappa$&\cite{kulin00}\\
\br
\end{tabular}
\end{indented}
\end{table}

Now consider a plasma where $\Gamma$ is fixed. From table \ref{apptab} we see that this will preserve the ratios $\nu_{\rme\rme}/\omega_p$ and $K/\omega_p$. Since $t_{\mathrm{exp}}$ depends on $m_\rmi$ the quantity $t_{\mathrm{exp}}\omega_p$ can be held constant regardless by adjusting the ion mass. Nevertheless, $\kappa$ depends explicitly on $N$ so a reduction in particle number means that $\kappa$ cannot stay constant. Thus it is impossible to preserve $\kappa$ and $\Gamma$ simultaneously. Furthermore, if $\kappa$ was fixed while scaling the particle number then $\Gamma$ must be adjusted thereby disrupting the scaling of the above collision rate to plasma frequency ratios. Therefore there is no way to create a scaling that keeps $\kappa$, $\Gamma$, $\nu_{\rme\rme}/\omega_p$, and $K/\omega_p$ all fixed.


\section*{References}
\begin{thebibliography}{10}
\bibitem{rolston99} Killian T C, Kulin S, Bergeson S D, Orozco L A, Orzel C and Rolston S L 1999 {\it Phys. Rev. Lett.} {\bf 83} 4776

\bibitem{pillet00} Robinson M P, Tolra B L, Noel M W, Gallagher T F and Pillet P 2000 {\it Phys. Rev. Lett.} {\bf 85} 4466

\bibitem{dutta01} Dutta S K, Feldbaum D, Walz-Flannigan A, Guest J R and Raithel G 2001 {\it Phys. Rev. Lett.} {\bf 86} 3993

\bibitem{kulin00} Kulin S, Killian T C, Bergeson S D and Rolston S L 2000 {\it Phys. Rev. Lett.} {\bf 85} 318

\bibitem{killian01} Killian T C, Lim M J, Kulin S, Dumke R, Bergeson S D and Rolston S L 2001 {\it Phys. Rev. Lett.} \textbf{86} 3759

\bibitem{murillo01} Murillo M S 2001 {\it Phys. Rev. Lett.} {\bf 87} 115003

\bibitem{ichimaru82} Ichimaru S 1982 {\it Rev. Mod. Phys} {\bf 54} 1017

\bibitem{bannasch12} Bannasch G, Castro J, McQuillen P, Pohl T and Killian T C 2012 {\it Phys. Rev. Lett.} {\bf 109} 185008

\bibitem{lyon13} Lyon M, Bergeson S D and Murillo M S 2013 {\it Phys. Rev. E} {\bf 87} 033101

\bibitem{lam65} Lam S H 1965 {\it Phys. Fluids} {\bf 8} 73

\bibitem{roberts04} Roberts J L, Fertig C D, Lim M J and Rolston S L 2004 {\it Phys. Rev. Lett.} {\bf 92} 253003

\bibitem{pillet05} Comparat D, Vogt T, Zahzam N, Mudrich M and Pillet P 2005 {\it Mon. Not. R. Astron. Soc.} {\bf 361} 1227

\bibitem{wilson13-2} Wilson T, Chen W-T and Roberts J L 2013 {\it Phys. Rev. A} {\bf 87} 013410

\bibitem{robicheaux02} Robicheaux F and Hanson J D 2002 {\it Phys. Rev. Lett.} {\bf 88} 055002

\bibitem{kuzmin02-1} Kuzmin S G and O'Neil T M 2002 {\it Phys. Rev. Lett.} {\bf 88} 065003

\bibitem{pohl04} Pohl T, Pattard T and Rost J M 2004 {\it Phys. Rev. A} {\bf 70} 033416

\bibitem{murillo06} Murillo M S 2006 {\it Phys. Rev. Lett.} {\bf 96} 165001

\bibitem{kuzmin02-2} Kuzmin S G and O'Neil T M 2002 {\it Phys. Plasmas} {\bf 9} 3743

\bibitem{christlieb06} Christlieb A J, Krasny R, Verboncoeur J P, Emhoff J W and Boyd I D 2006 {\it IEEE Trans. Plasma Sci.} {\bf 34} 149

\bibitem{jeon08} Jeon B, Kress J D, Collins L A and Gr\o nbech-Jensen N 2008 {\it Comput. Phys. Commun.} {\bf 178} 272

\bibitem{bannasch11} Bannasch G and Pohl T 2011 {\it Phys. Rev. A} {\bf 84} 052710

\bibitem{barnes86} Barnes J and Hut P 1986 {\it Nature} {\bf 324} 446

\bibitem{greengard87} Greengard L and Rokhlin V 1987 {\it J. Comput. Phys.} {\bf 73} 325

\bibitem{wilson13-1} Wilson T, Chen W-T and Roberts J 2013 {\it Phys. Plasmas} {\bf 20} 073503

\bibitem{robicheaux03} Robicheaux F and Hanson J D 2003 {\it Phys. Plasmas} {\bf 10} 2217

\bibitem{collins08} Vrinceanu D, Balaraman G S and Collins L A 2008 {\it J. Phys. A: Math Theor.} {\bf 41} 425501

\bibitem{rolston10} Twedt K A and Rolston S L 2010 {\it Phys. Plasmas} {\bf 17} 082101

\bibitem{spitzer} Spitzer L 1962 {\it Physics of Fully Ionized Gases} (Mineola, NY: Dover Publications)

\bibitem{raithel08} Choi J-H, Knuffman B, Zhang X H, Povilus A P and Raithel G 2008 {\it Phys. Rev. Lett.} {\bf 100} 175002

\bibitem{arfken} Arfken G B, Weber H J and Harris F E 2013 {\it Mathematical Methods for Physicists} (Waltham, MA: Elsevier)

\bibitem{kubo} Kubo R, Ichimura H, Usui T and Hashitsume N 1990 {\it Statistical Mechanics} (Amsterdam: North-Holland)

\bibitem{witte14} Witte C and Roberts J L 2014 {\it Phys. Plasmas} {\bf 21} 103513

\end{thebibliography}
\end{document}